\documentclass[aps, twocolumn, prl, groupedaddress, showpacs, floatfix]{revtex4-1}
\usepackage[dvips]{graphicx}
\usepackage{amsmath}
\usepackage{amsfonts}
\usepackage{color}
\usepackage{times}

\DeclareMathOperator{\Rep}{Re}

\begin{document}
\renewcommand{\topfraction}{0.98}	        
\renewcommand{\bottomfraction}{0.98}        
\setcounter{topnumber}{3}
\setcounter{bottomnumber}{3}
\setcounter{totalnumber}{4}                 
\setcounter{dbltopnumber}{4}                
\renewcommand{\dbltopfraction}{0.98}        
\renewcommand{\textfraction}{0.05}	        
\renewcommand{\floatpagefraction}{0.5}	    
\renewcommand{\dblfloatpagefraction}{0.5}	
\sloppy 
\newcommand{\beq}{\begin{equation}}
\newcommand{\eeq}{\end{equation}}
\newtheorem{rmk}{Remark}
\newcommand{\divg}{\mbox{\rm{div}}\,}
\newcommand{\Divg}{\mbox{\rm{Div}}\,}
\newcommand{\D}  {\displaystyle}
\newcommand{\DS} {\displaystyle}
\def\sca   #1{\mbox{\rm{#1}}{}}
\def\mat   #1{\mbox{\bf #1}{}}
\def\vec   #1{\mbox{\boldmath $#1$}{}}
\def\scas  #1{\mbox{{\scriptsize{${\rm{#1}}$}}}{}}
\def\scaf  #1{\mbox{{\tiny{${\rm{#1}}$}}}{}}
\def\vecs  #1{\mbox{\boldmath{\scriptsize{$#1$}}}{}}
\def\tens  #1{\mbox{\boldmath{\scriptsize{$#1$}}}{}}
\def\ten   #1{\mbox{\boldmath $#1$}{}}
\newcommand{\C}{\mathbb{C}}
\newcommand{\abs}[1]{\left|#1\right|}
\renewcommand{\Re}{\Rep}
\newcommand{\Lb}{\mathcal{R}}
\newcommand{\sset}[1]{\left\{#1\right\}}

\title{Neuronal Oscillations on Evolving Networks:\\ Dynamics, Damage, Degradation, Decline,  Dementia, and Death}
\author{Alain Goriely$^1$, Ellen Kuhl$^2$, and Christian Bick$^{3,1,4}$}
\affiliation{
\it $^1$Mathematical Institute, University of Oxford, Oxford, OX2 6GG, UK\\
\it $^2$Living Matter Laboratory, Stanford University, Stanford, CA 94305, USA\\
\it $^3$Department of Mathematics, University of Exeter, Exeter, EX4 4QF, UK\\
\it $^4$Institute for Advanced Study, Technische Universit\"at M\"unchen, 85748~Garching, Germany}
\begin{abstract} 
\noindent 
Neurodegenerative diseases, such as Alzheimer's or Parkinson's disease, show characteristic degradation of structural brain networks. This degradation eventually leads to changes in the network dynamics and  degradation of cognitive functions. 
Here, we model the progression in terms of coupled physical processes: The accumulation of toxic proteins, given by a nonlinear reaction-diffusion transport process, yields an evolving brain connectome characterized by weighted edges on which a neuronal-mass model evolves. The progression of the brain functions can be tested by simulating  the resting-state activity on the evolving brain network. We show that while the evolution of edge weights plays a minor role in the overall progression of the disease, dynamic biomarkers predict a transition over a period of 10 years associated with strong cognitive decline.
\end{abstract}
\maketitle

\noindent\textit{Introduction.---}%
Neurodegenerative diseases are not only major health and societal problems~\cite{El-Hayek2019}, they are also formidable scientific challenges. Their neuropathology, characterized by diseased brain tissue and  cortical atrophy, is linked to the accumulation of toxic proteins. These \emph{structural modifications} change the way neurons interact~\cite{Palop2010} and lead to cognitive decline and neurobehavioral symptoms~\cite{Koch2015}. Specifically, axonal death  has a direct effect on the collective \emph{brain network dynamics}, including synchronization~\cite{Uhlhaas2006}, dynamics~\cite{Zimmermann2018,Budzinski2019} and  connectivity~\cite{wang2007altered}. There are three interconnected  physical and cognitive processes at work: disease progression through the brain, structural damage created by the disease, and dynamic changes from damage with the associated functional loss.

Here, we build a model that  predict both the spatio-temporal evolution of the disease but also how it affects basic cognitive functions. Our approach combines dynamics on multiple temporal scales (years for the disease and seconds for the resting-state dynamics) with multiphysics at various levels (transport, aggregation, damage, and oscillations). Specifically, we look at interacting dynamical processes on an evolving network structure: Disease progression changes the structural properties of the brain connectome, which results in changes to characteristic dynamics of the brain network dynamics. To probe cognitive functions of a given connectome we simulate whole-brain resting states against functional magnetic resonance imaging (fMRI) resting-state data~\cite{Breakspear2017}. Since Gamma activity emerge in neural populations~\cite{Fisahn1998} and is related to both hippocampal memory formation~\cite{chrobak1998gamma} and  Alzheimer's disease~\cite{iaccarino2016gamma,mcdermott2018gamma}, we use a minimal neural-mass model with intrinsic frequency in the Gamma range (defined by $\Gamma = [30, 100]\ \textrm{Hz}$)~\cite{deco2009key}.

\textit{Disease progression.---}%
We follow the prion-like paradigm~\cite{jucker2011,walker2015neurodegenerative,goedert2015alzheimer}, which proposes that degeneration is caused by the invasion and conformational autocatalytic conversion of misfolded proteins transported along axonal pathways.~\cite{Prusiner1998}.
The prion-like idea has served as an important unifying concept through which many features  can be understood such as staging~\cite{braak1991neuropathological},  biomarker evolution~\cite{jack2013biomarker}, and neural atrophy. Basic models based in this notion recover most of these observations~\cite{Weickenmeier2018,Weickenmeier2019}.
The basic features of these diseases can be obtained by restricting all physical quantities on the structural connectome, a brain network representing the connections between different regions of interest~\cite{raj2012network,iturria2014epidemic,henderson2019spread}.  Our model~\cite{sveva,fornari2020spatially}, with parameters taken from~\cite{kundel2018measurement}, combines network diffusion encoded by a graph Laplacian and a reaction term characterizing the population amplification  due to the conversion of healthy proteins.

\begin{figure}
\centerline{\includegraphics[width=0.9\linewidth]{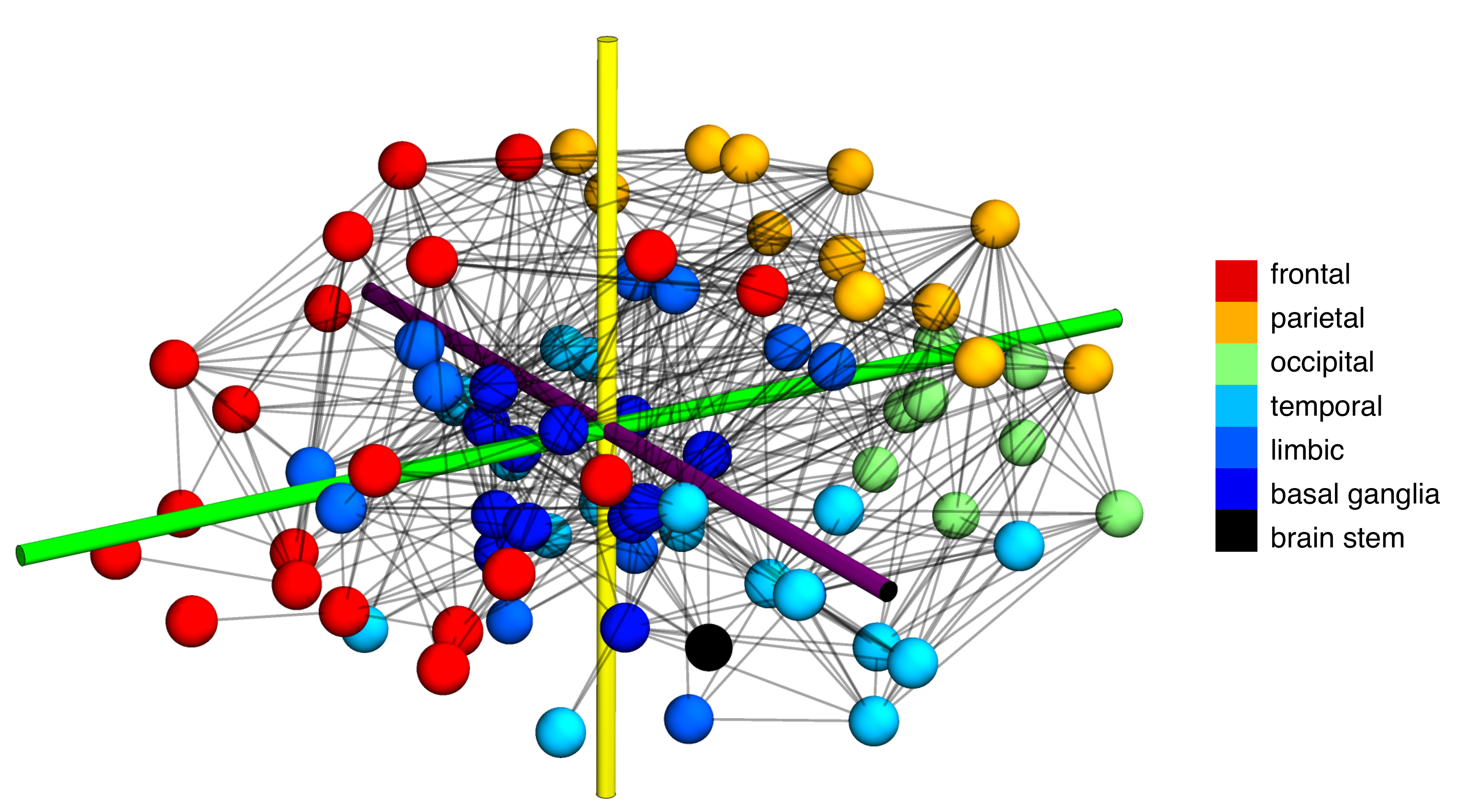}}
\caption{\label{Fig-adjacency}The  connectome ${\mathcal{G}_0}$  with $N=83$ nodes and $M=1654$ edges (592 shown).}
\end{figure}
We model the connectome as an evolving undirected weighted graph~${\mathcal{G}_T}$ at time~$T\geq0$. The initial graph~${\mathcal{G}_0}$ is extracted from  from the tractography of diffusion tensor magnetic resonance images of 418 healthy subjects of the Budapest Reference Connectome v3.0~\cite{mcnab2013human,szalkai2017parameterizable}.
Each node~$k$ is associated with a particular brain region~$\Lb_s$, $s=\sset{1,\dotsc,7}$, corresponding to the frontal, parietal, temporal, occipital lobes, the limbic area, the basal ganglia, and the brain stem; cf.~Fig.~\ref{Fig-adjacency}.
The graph has~$M$ edges with weight $w_{kj}(T)$,  between nodes~$k$ and~$j$, defined as the ratio of the number of fibers between the nodes and the mean fiber length, which leads to the symmetric matrix $\mathbf{W}=(w_{kj})$, $k,j\in\{1,\dots,N\}$ and  the weighted graph Laplacian $\mathbf{L}=\rho(\mathbf{D}-\mathbf{W})$ where $\mathbf{D}=\text{diag}($\mbox{$\sum_{j=1}^{N}$}$ w_{kj})$ and~$\rho$ is an overall velocity constant defining the time-scale of transport. 

Assuming  the concentration of healthy proteins remains mostly unchanged \cite{Weickenmeier2019}, the protein concentration~$c_k(t)$ at node~$k$ obeys a network discretization of a Fisher--KPP equation~\cite{sveva} given by
\begin{align}\label{c}
\dot c_k &= -\sum_{j=1}^N L_{kj} c_j+\alpha c_k (1-c_k),&  k&=1,\dotsc,N,
\end{align}
where~$\alpha$ characterizes the conversion from healthy to toxic proteins. Since~$L$ depends on the  weights~$\mathbf{W}$, we  require an equation for the evolution of the weights in time.

\textit{Network damage and evolution.---}%
The accumulation of toxic proteins  influences the network properties due to its effect on synapses, plasticity, and eventual cell death~\cite{Collinge1994,Palop2010,Palop2016}.
We quantify the damage at each node by a variable $q_k\in[0,1]$  ($0$ healthy, $1$ maximal damage),
for which we assume a first-order rate model:
\begin{align}\label{q}
\dot q_k &=\beta c_k (1-q_k),\quad q_k(0)=0,\quad & k&=1,\dotsc, N,
\end{align}
where~$\beta$ characterizes the protein toxicity. Since transport away from a node depends on the node's health, the damage at a node affects the connectivity to other nodes. We assume that the relative change of an edge weight depend on the damages at the nodes with a rate~$\gamma$ according to
\begin{align}\label{w}
\dot w_{kj} &= -\gamma w_{kj} (q_k+q_j), & k,j&=1,\dotsc, N,
\end{align}
and the systems~(\ref{c}--\ref{w}) form a closed system of $2N+M$ ordinary differential equations. Initially, the concentration of toxic proteins vanishes at all nodes except at seeding nodes that is disease-dependent. For illustrative purpose, we use the propagation of tau proteins as the main source of toxic protein and  seed the system in the entorhinal region \cite{de2012propagation} by setting $c_{k}(0)=0$ for all~$k$ except for $c_{26}(0)=c_{68}(0)=0.025$. Initially,  the connectome is healthy, with $q_k(0)=0$ for all~$k$, and  the weights~$w_{kj}(0)$ are given by~$\mathcal{G}_0$.
\begin{figure}
\centerline{\includegraphics[width=0.9\linewidth]{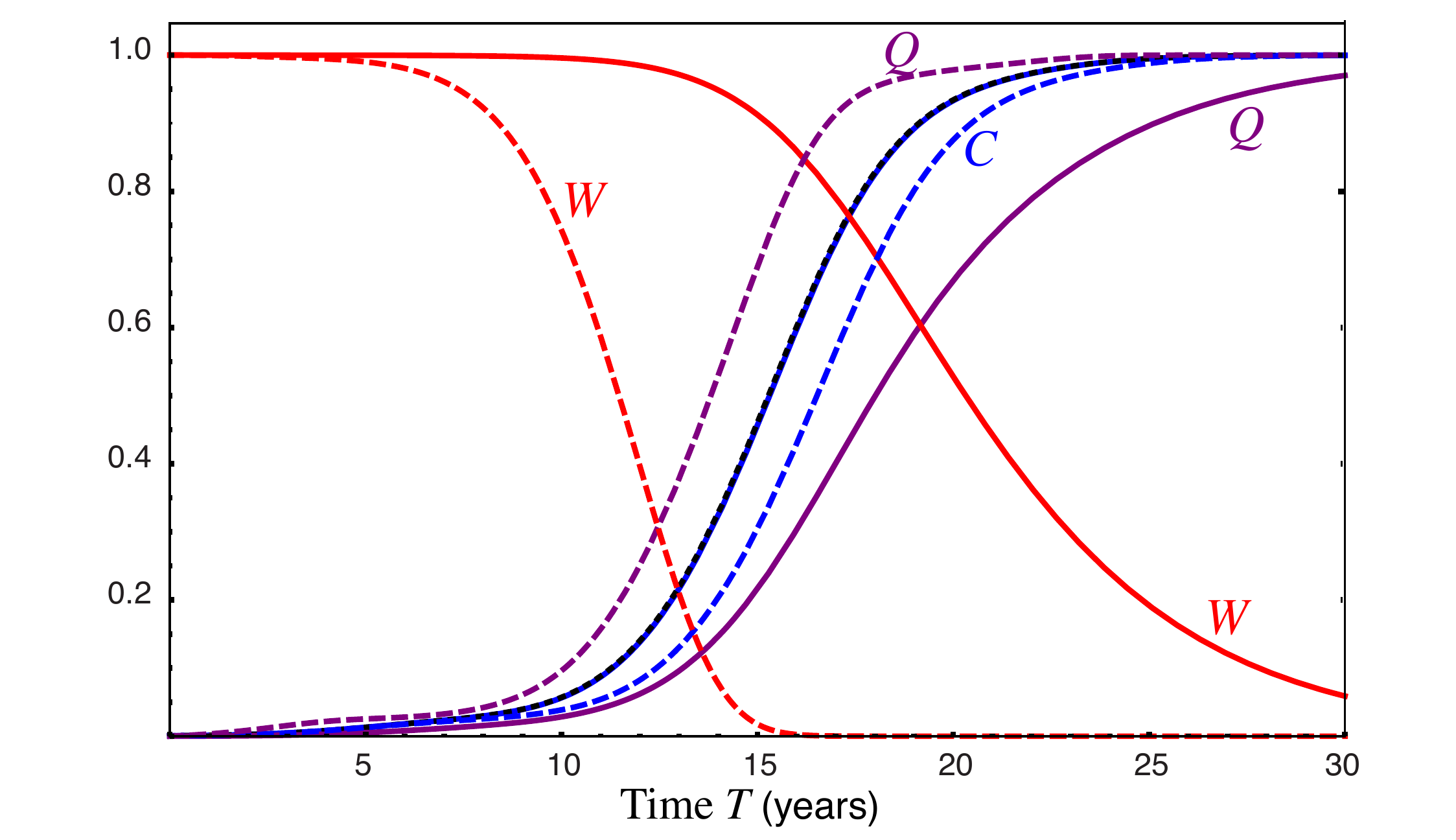}}
\caption{\label{Fig-damage} Evolution of averaged toxic concentration and  damage. The black dotted curve (superimposed with the blue solid curve) is the average concentration  in the absence of damage ($\beta=\gamma=0$) whereas the solid curve is the case of severe damage ($\beta=(1/4)$/year, $\gamma=(1/8)$/year) leading to a reduction of the connection weight of 50\% after 20 years. Even for unrealistic values of the parameters ($\beta=4,\gamma=2$-dashed curve) with a reduction of 99\% of all weights after 15 years, the delay in the evolution of the concentration is only a year ($\alpha=(3/4)$/year, $\rho=1/100$ mm/year in all simulations). }
\end{figure}

\textit{Damage does not slow down disease progression.---}%
Before considering the resting-state dynamics on an evolving connectome, we study the effect of damage on the propagation of the disease. To quantify disease progression, we compute three key {structural biomarkers} evaluated at a sampled time $T$: (a)~the {average concentration}
$C(T)=\frac{1}{N}\sum_{j=1}^{N} c_j(T)$,
(b)~the {average damage} 
$Q(T)=\frac{1}{N}\sum_{j=1}^{N} q_j(T)$,
and (c)~the {scaled average connection weight}
$W(T)={\|\mathbf{W}(T)\|}/{\|\mathbf{W}(0)\|}$, 
where $\|{\mathbf{W}}\|=N^{-2}\sum_{k,j=1}^{N} w_{kj}$.
We also compute the average damage~$Q_s$ in region~$s$ by only summing over indices in~$\Lb_s$ and normalizing accordingly. 

While one may expect a slowing down of disease progression as the transport network is affected, the actual overall effect is negligible as shown in Fig.~\ref{Fig-damage}. We first compare $C$ in the absence of damage (solid curves), with the case of severe damage over a period of 30~years (dotted curve) and see a negligible difference. Even for unrealistic values (dashed curves) leading to the destruction of the network within a period of 15~years, the delay in invasion is only about a year. Hence, network damage does not slow down significantly the invasion of the disease even in extreme cases. Indeed, the reduction of diffusion associated with damage mostly affects regions where the concentration is high. In these regions, most of the tissue is already damaged and only very little transport is taking place. In terms of front dynamics, the front velocity in  Fisher--KPP  depends entirely on the asymptotic zero state. Once nodes have been seeded, the local increase in concentration takes place even in the absence of a network. This nonlinear effect due to the conversion of healthy to toxic agents is fundamentally different than the diffusion process where  the toxic protein must be carried from its source.

\begin{figure}
\centerline{\includegraphics[width=0.9\linewidth]{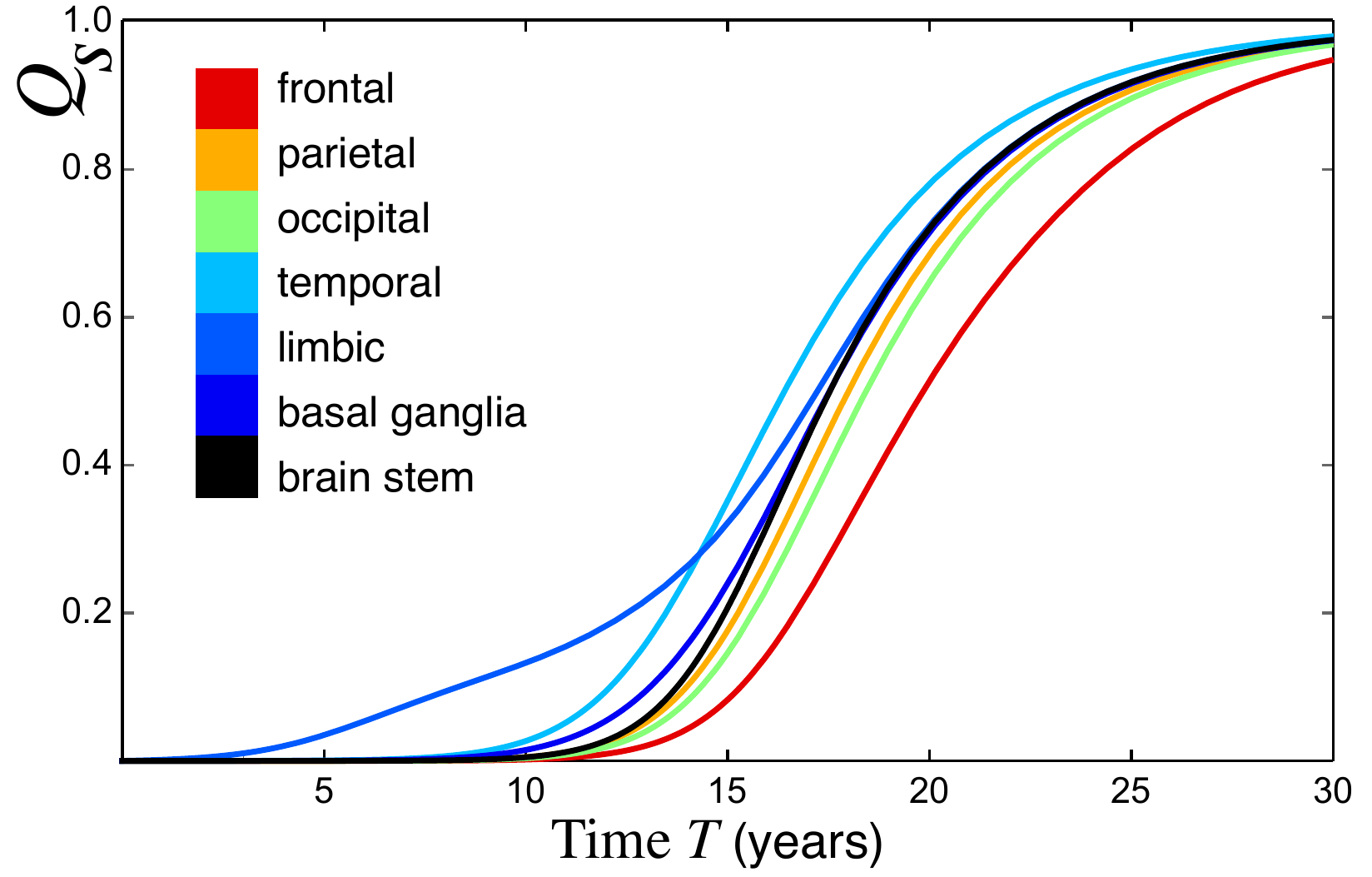}}
\caption{\label{Fig-damage-lobe} Evolution of  damage in different brain regions. Physical damage first manifests itself in the limbic region then moved to the temporal lobe, the basal ganglia, and the parietal and occipital lobes before invading all cortical areas. }
\end{figure}

Disease staging can be obtained by computing damage in each region (Fig.~\ref{Fig-damage-lobe})  \cite{kesslak1991quantification}. Physical damage first appears in the limbic region where the disease originates but moves rapidly to the temporal and parietal lobes. This early invasion can be understood by looking at the topological properties of the linearized system \cite{fornari2020spatially} (that can be solved explicitly~\cite{go01} as shown in the Supplemental Material). Note that increase in damage in the limbic region is slower than in other regions and that by year~13 the total damage for example in the temporal lobe is larger than in the limbic region. Eventually, the disease invades all cortical areas. 

\textit{Resting state brain dynamics.---}%
To test the declining cognitive functions of the brain, we focus on resting-state brain dynamics. The time scales involved in the process are of the order of months for the disease and of the order of seconds for the rest-state activity. Therefore, the disease dynamics is quasi-stationary and at time $t=T$ we consider the connectome~$\mathcal{G}_{T}$ to be constant when probing resting-state activity.
As a proof of principle, we consider a simple neural-mass model on each node representing large interacting excitatory and inhibitory neural populations to approximate a Wilson--Cowan type model~\cite{deco2009key,Deco2017}.
In the absence of coupling, the intrinsic node dynamics are given by a supercritical Hopf bifurcation.
The state of node~$k$ is given by~$z_k\in\mathbb{C}$, and, apart from an offset, the real part of~$z_k$ encodes the {activity of the excitatory population} and the imaginary part the {activity of the inhibitory population}.
For a given network~$\mathcal{G}_{T}$ with associated weights~$\mathbf{W} = \mathbf{W}(T)$ the neural populations are coupled through the amplitude of the excitatory population and modulated by a sigmoidal function $S(x) = 1/(1+\exp(-x))$ through the delay differential equation
\begin{align}\label{eq:NeuroDyn}
\dot z_k = F(z_k) + \kappa S\!\left(\Re\left(\sum_{j=1}^N w_{kj}z_j(t-\tau_{kj})\right)\right),
\end{align}
where $F(z_{k}) = z\big(\lambda+i\omega_{k} - \left|z_{k}^2\right|\big)$ with decay $\lambda = -0.01$, intrinsic frequencies $\omega_k=\omega+\delta_k=40\mathrm{Hz}+\delta_k$ whose deviations~$\delta_k$ are sampled from a normal distribution (mean zero, variance~$0.1\mathrm{Hz}$), coupling gain $\kappa=10$, and delays~$\tau_{kj}$ proportional to the distance between node~$k$ and~$j$ from the connectome data with transmission speed of~$1.5\mathrm{m}/\mathrm{s}$ (discretized to have a maximum of 40 distinct delays); these model parameters were chosen to approximate the neural-mass model in~\cite{deco2009key} validated against resting-state fMRI. For the initial graph~$\mathcal{G}_0$, we observe collective oscillations. In the absence of coupling, all amplitudes decay exponentially with a frequency of around 40Hz.

\textit{Global cognitive decline after physical damage---}%
As indicators of cognitive processes, 
we consider three {dynamic biomarkers} obtained from~$t_\text{sim} = 10\mathrm{s}$ of resting-state dynamics~\eqref{eq:NeuroDyn}: 
(d)~the {overall power in the Gamma-range}
$P(T) =\int_\Gamma \textrm{PSD}(\langle z\rangle)(\Omega)\,\textrm{d}\Omega$,
where $\textrm{PSD}$ is the power spectral density of the signal $\langle z\rangle = \frac{1}{N}\sum_{j=1}^{N} \Re(z_{j})$,
(e)~the {average oscillatory activity} 
$A(T) = {N}^{-1}\sum_{j=1}^N{t_\text{sim}^{-1}}\int_{0}^{t_\text{sim}} |z_j(t)|\,\textrm{d}t$,
and
(f)~the {metastability index} 
$B(T) ={N}^{-1}\sum_{j=1}^N \sigma^2_t(|z_j(t)|)$,
where $\sigma^2_t$ is the variance of the signal over the time interval $[0, t_\text{sim}]$.
We define the corresponding measures~$P_{s}, A_{s}, B_{s}$ for~$\Lb_s$ by summing over the corresponding nodes and normalizing accordingly. These dynamic biomarkers have been associated with cognitive processes and  neurodegenerative diseases~\cite{Uhlhaas2006}. The average amplitude is a measure of the general activity  and the metastability index is associated with information processing~\cite{rabinovich2012information,Tognoli2014,Heitmann2018,shanahan2010metastable}.

To evaluate how the disease affects the dynamics, we solved~\eqref{eq:NeuroDyn} on the evolving brain connectome for different times~$T$; see Supplemental Material for details.
Fig.~\ref{fig:Evol} shows the mean and standard deviations of the dynamical biomarkers scaled with respect to the healthy response at~$T=0$. We see that all dynamical indicators remain fairly unchanged up to year~20 as the disease progresses. At that time, the brain has already suffered significant physical damage even if this damage cannot be easily assessed from the dynamics (see Fig.~\ref{Fig-damage}). It suggests that the brain state is structurally stable against damages for an extended time. However, after 20 years, the dynamics undergoes a clear transition when nodes are unable to sustain oscillatory activities.
\begin{figure}
\centerline{\includegraphics[width=0.9\linewidth]{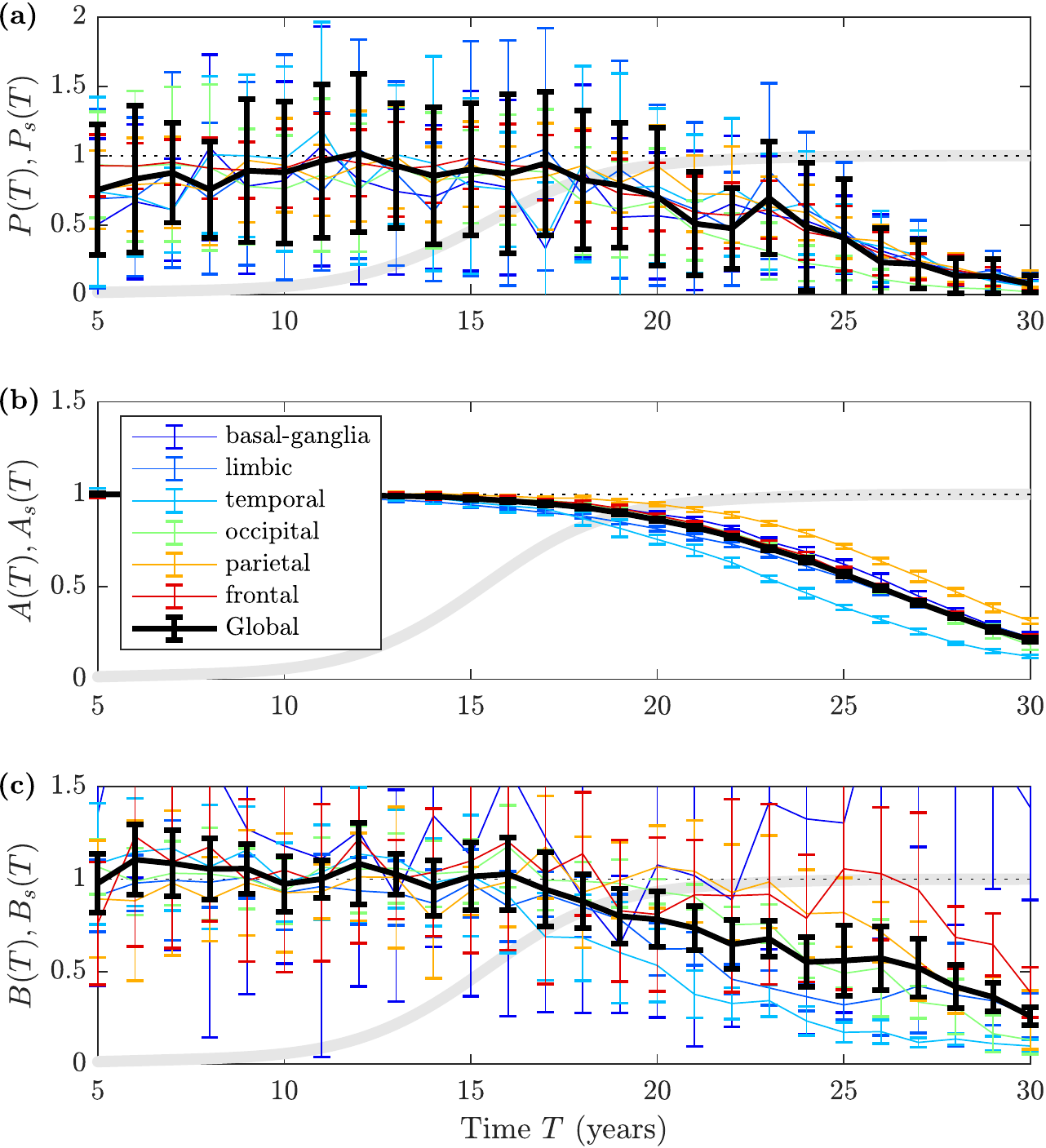}}
\caption{\label{fig:Evol}%
Dynamic biomarkers are stable up to year 10 followed by a rapid decline from year 16 onward.
(a)~power in the $\Gamma$ band; (b)~oscillation mean amplitude; there are no oscillations in the absence of network coupling. (c)~mean variability which indicates non-stationary (and potentially metastable) brain dynamics.
Mean and standard deviation for 12 realizations with different intrinsic frequencies and initial conditions (solid gray line is~$C(T)$ for comparison).}
\end{figure}

The local dynamical biomarkers show a more differentiated picture of how the spreading disease changes the dynamics. The transition of the global dynamics is preceded by a decay of oscillatory activity in the temporal lobe. In terms of the oscillatory activity, the markers for all other regions are close to the global mean (Fig.~\ref{fig:Evol}b). 
Given that damage first accumulates in the limbic system (Fig.~\ref{Fig-damage-lobe}), this observation may be counterintuitive. However, it indicates that both speed of damage accumulation and network structure determine the critical threshold for the decay of oscillatory dynamics: The region where one first sees significant structural damage can be distinct from the region that first undergoes a dynamical transition. This is consistent with experimental findings that indicate that temporal lobe activity is a precursor for the disease onset~\cite{Beason-Held2013}.

\textit{Adaptation slows cognitive decline.---}%
Connections between neurons are typically not static but adjust in response to the dynamics~\cite{Sjostrom2008,Markram2011}. 
Thus, in addition to damage-induced changes to the network, the neural dynamics itself affects the network to maintain homeostasis~\cite{Frere2018}. To capture this effect, we implement a minimal model of homeostatic adaptation that aims to keep the mean connectome coupling  constant over time. For an adaptation parameter~$\xi\in[0, 1]$, we define the scaled matrix with ${\overline{\mathbf{W}}}(0) =\mathbf{W}(0)$ and
\begin{equation}\label{eq:Homeostasis}
{\overline{\mathbf{W}}}(T) = \left((1-\xi)+\xi\left(\frac{\|{{\overline{\mathbf{W}}}(T-1)}\|}{\|{{\mathbf{W}(T)}}\|}\right)\right)\mathbf{W}(T)
\end{equation}
for $T = 1,2,\dotsc$.
This homeostatic adaptation creates a new weighted connectome for the resting-state dynamics~\eqref{eq:NeuroDyn} for every year~$T$ of disease progression. Through~\eqref{eq:Homeostasis}, homeostasis acts by globally rescaling the coupling weights. For $\xi=0$, there is no homeostatic adaptation and the network structure changes solely through the disease progression.
For $\xi=1$, there is complete homeostatic adaptation in the sense that we have a constant mean coupling weight $\|{{\overline{\mathbf{W}}}(T)}\|=1$ for all~$T$.
This does not imply $\overline{\mathbf{W}}(T)=\overline{\mathbf{W}}(0)$. Rather, due to the global nature of the homeostatic adaptation, disease-induced changes in coupling strength in one brain region will yield hyper-excitability in another brain region to balance the overall decay in coupling strength.

\begin{figure}
\centerline{\includegraphics[width=0.9\linewidth]{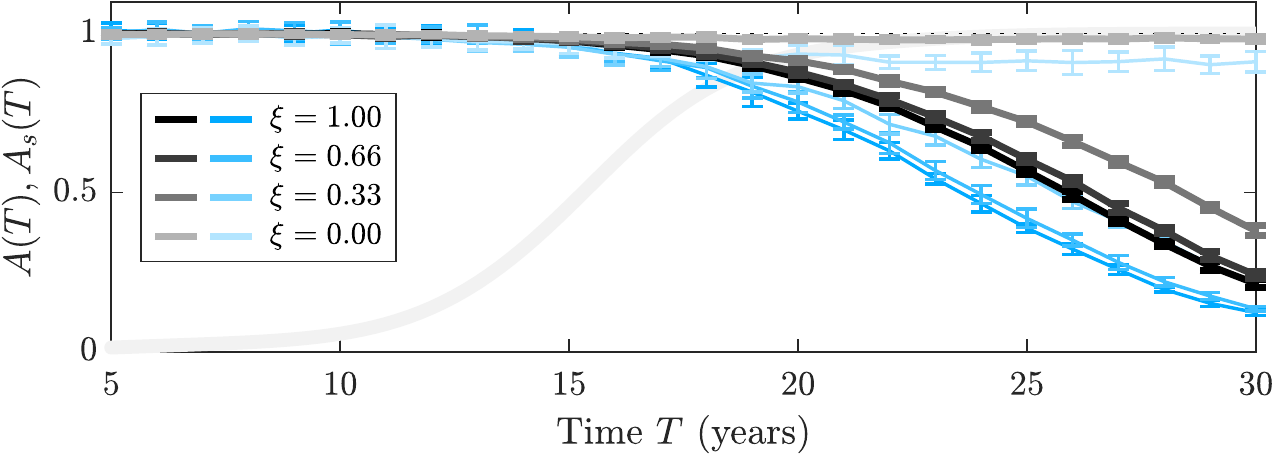}}
\caption{\label{fig:Homeo}
Oscillation amplitude changes with the homeostasis parameter. Increasing homeostasis does not lead to a shift in the onset of decline but only changes the shape of the decline. The global mean oscillation amplitude (gray lines) shows how the slope varies for different homeostasis parameters; in the extreme case $\xi=1$, the oscillation amplitude remains constant. By contrast, the mean amplitude in the temporal lobe (blue lines) decreases in any case. For comparison, $C(T)$ is given by the solid gray line.
}
\end{figure}

Homeostatic adaptation modulates the disease progression. We simulated the resting-state dynamics~\eqref{eq:NeuroDyn}  subject to the homeostatic adaptation~\eqref{eq:Homeostasis}. Fig.~\ref{fig:Homeo} shows the average amplitude~$A(T)$ as disease progresses for different values of the adaptation parameter~$\xi$. An increase in homeostasis does not change the actual onset of loss of oscillatory activity (around year 13) but yields a slow-down of the degeneration. However, assuming that cognitive decline sets in once a certain threshold is reached, the slow-down of disease progression will alter the onset of the overall transition.
Fig.~\ref{fig:Homeo} also shows mean oscillation amplitude for the temporal lobe. The temporal lobe shows a decay in oscillatory activity for all values of the adaptation parameters preceding the overall decay of oscillatory activity. This happens even for full adaptation, $\xi=1$, where the overall oscillatory activity stays constant in the time-window considered here. This decline implies that other brain areas have to be upregulated in order to keep the overall activity almost constant. However, it also means that for our simple model of adaptation, the variation of oscillation amplitude  is a precursor for the overall transition of the dynamics {independently of the adaptation parameter}.

\textit{Discussion.---}%
Compared to previous studies that focused on either static properties of the declining network~\cite{stam2006small}, synchronization \cite{Uhlhaas2006} or activity-based decline~\cite{de2012activity}, we focused on the importance of the underlying interacting physical processes. The physical damage of the brain and neural dynamics---two independent yet coupled processes---interact on the same connectome. 
Neurodegeneration, modeled as an invasion process due to the accumulation of toxic proteins, provides a natural evolution of the connectome on long time scales that can be probed dynamically on short time scales. Our observations are compatible with the activity-dependent spreading hypothesis~\cite{de2012activity}: toxic proteins  will spread predominantly along highly connected nodes which also show high activity due to the amount of input they receive. 

While the actual brain connectome is not an undirected graph, there is very little information on  directionality for the human brain. However,  an analysis of a mesocale mouse connectome~\cite{oh2014mesoscale} that has been used for toxic protein diffusion~\cite{henderson2019glucocerebrosidase,henderson2019spread} reveals that asymmetry delays significantly the onset of the disease but preserves its main characteristics (see Supplemental Material).

Our setup provides a unified framework that combines the biophysics of disease spreading with whole-brain dynamics to give mechanistic insights into the dynamics of neurodegenerative diseases and its associated cognitive decline. First, we showed that damage does not slow the disease propagation as damage is delayed with respect to seeding. Very low level of toxic proteins diffuse and seed new regions. Then, even in the absence of transport, there is a local autocatalytic increase of toxic protein. Second, we gained insight into the dynamical transitions appearing in some brain regions and confirms the prediction~\cite{Beason-Held2013} that the temporal lobe is one of the first to see alteration in brain dynamics, hence showing cognitive deficiencies related to that region for Alzheimer's disease as shown in Fig.~\ref{fig:Evol}. Third, our results elucidate the interplay between network adaptation and spreading. We found that incorporating a simple model of global homeostasis does not change the onset of dynamical changes, but how fast they evolve as the disease progresses. Interestingly, we saw that, in the particular case of Alzheimer disease, a decline of oscillations in the temporal lobe is a universal indicator independent of the adaptation parameter.

Further insights are needed on how the multiphysics of disease propagation interact with brain network dynamics to identify reliable noninvasive biomarkers to assess disease progression as early as possible and develop an integrated approach for treatment. 
Other models for whole-brain dynamics are available that focus on different features of brain dynamics~\cite{Breakspear2017} and relate to microscopic neural properties~\cite{Bick2018c}.
When combined with data from a comprehensive longitudinal study and suitable dynamical models, our approach will be able to explore new treatment approaches and intervention scenarios on the combined level of network structure and dynamics~\cite{DeHaan2017,Palop2010,Canter2016}.

\begin{acknowledgments}
This work was supported by the Engineering and Physical Sciences Research Council grant EP/R020205/1 to AG and by the National Science Foundation grant CMMI 1727268 to EK.
\end{acknowledgments}

\bibliographystyle{apsrev4-1}
\def\urlprefix{}
\def\url#1{}
%

\end{document}